\documentclass[a4paper]{jpconf}
\usepackage{graphicx}
\usepackage{subfigure}
\usepackage{bm}
\usepackage{amsmath, amsfonts}
\begin{document}
\title{Semi-supervised anomaly detection -- towards model-independent searches of new physics}

\author{Mikael Kuusela$^{1,2}$, Tommi Vatanen$^{1,2}$, Eric Malmi$^{1,2}$, Tapani Raiko$^{2}$, Timo Aaltonen$^{1}$ and Yoshikazu Nagai$^{3}$}

\address{$^1$ Helsinki Institute of Physics, PO Box 64, FI-00014 University of Helsinki, Finland}
\address{$^2$ Aalto University, Department of Information and Computer Science, PO Box 15400, FI-00076 Aalto, Finland}
\address{$^3$ University of Tsukuba, 1-1-1 Tennodai, Tsukuba, Ibaraki 305-8577, Japan}

\ead{\{mikael.kuusela,tommi.vatanen,eric.malmi\}@aalto.fi}

\begin{abstract}
Most classification algorithms used in high energy physics fall under the category of supervised machine learning. Such methods require a training set containing both signal and background events and are prone to classification errors should this training data be systematically inaccurate for example due to the assumed MC model. To complement such model-dependent searches, we propose an algorithm based on semi-supervised anomaly detection techniques, which does not require a MC training sample for the signal data. We first model the background using a multivariate Gaussian mixture model. We then search for deviations from this model by fitting to the observations a mixture of the background model and a number of additional Gaussians. This allows us to perform pattern recognition of any anomalous excess over the background. We show by a comparison to neural network classifiers that such an approach is a lot more robust against misspecification of the signal MC than supervised classification. In cases where there is an unexpected signal, a neural network might fail to correctly identify it, while anomaly detection does not suffer from such a limitation. On the other hand, when there are no systematic errors in the training data, both methods perform comparably.
\end{abstract}

\section{Introduction}

Machine learning techniques have been extensively used in high energy physics data analysis over the past few decades \cite{Bhat2011}. Especially supervised, multivariate classification algorithms, such as neural networks and boosted decision trees, have become commonplace in state-of-the-art physics analyses and have proven to be invaluable tools in increasing the signal-to-background ratio in searches of tiny signals of new physics. Such methods work under the assumption that there exists labeled data sets of signal and background events which can be used to train the classification algorithm. These training samples are usually obtained from Monte Carlo (MC) simulation of the new physics process and the corresponding Standard Model background.

Supervised machine learning algorithms are the tools of model-dependent new physics searches. In the beginning of the analysis, one decides to focus on a particular variant of a certain new physics model. This could be for example a certain SUSY parametrization. Events corresponding to this choice are then generated using a MC generator followed by training of a classification algorithm to optimize the yield of such events. In the best case scenario, applying this classifier to real observed data would produce a statistically significant excess over the Standard Model background and result in a discovery of new beyond the Standard Model physics.

Unfortunately, this process could go wrong on several different levels. Firstly, there are no guarantees that nature actually obeys any of the existing theoretical new physics models. The solution to the existing problems with the Standard Model could be something that no one has even thought of yet. Even if one of the existing theories is the right way forward, such models often have a large amount of free parameters. Exploring the whole high-dimensional parameter space one combination of values at a time could be a very laborious, time-consuming and error-prone process. Lastly, MC simulation of such processes often requires simplifications and approximations and could be a major source of systematic errors in its own right.

\begin{figure}[tb]
\centering
\subfigure{
\includegraphics[width=6cm]{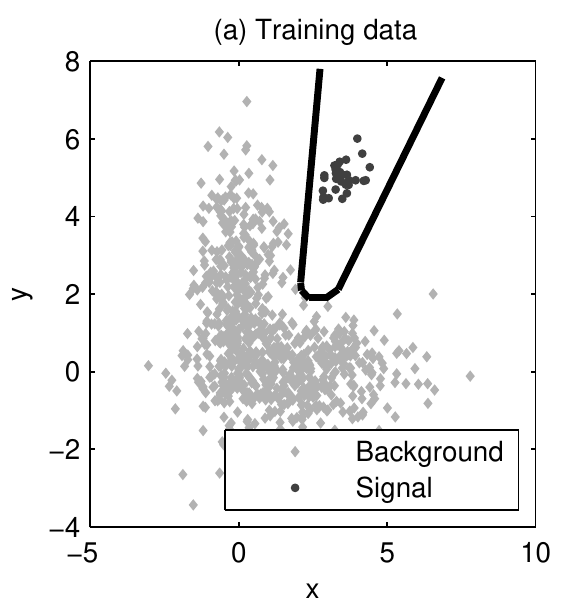}}
\hspace{1cm}
\subfigure{
\includegraphics[width=6cm]{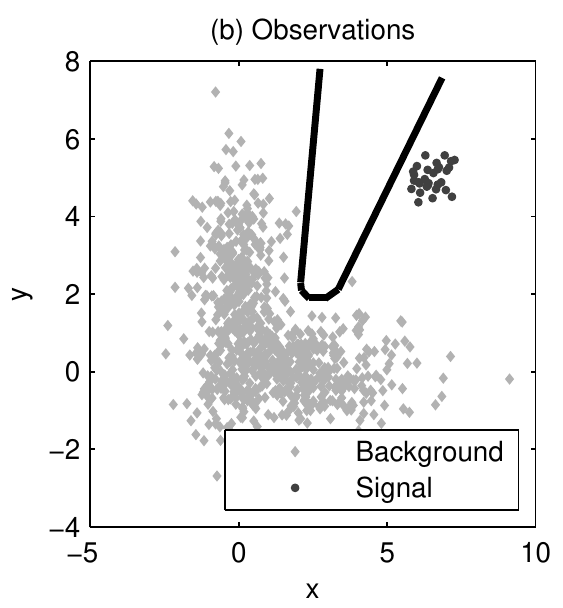}}
\caption{Demonstration of the limitations of model-dependent searches of new physics. When a neural network is trained with the data of the left figure, the classification decision boundary is given by the black curve. But if the training data is systematically inaccurate, a new physics signal could be completely missed in the analysis. For example all the observations of the right figure would be classified as background.}
\label{fig:nnDemo}
\end{figure}

This kind of uncertainties and systematic errors are not well-tolerated by supervised classification algorithms. Figure \ref{fig:nnDemo} illustrates the problem. When a neural network was trained using the signal and background data of Figure \ref{fig:nnDemo}a, we obtained the black decision boundary to separate signal events from background. But what if due to one of the above mentioned problems, the real new physics signal looked like the one in Figure \ref{fig:nnDemo}b? The algorithm trained with wrong type of training patterns would completely miss such a signal. The signal events would all be regarded as background even though the data clearly contains a signature of an interesting anomalous process.

While it is possible to alleviate this type of problems up to some extent by tweaking the training data and the classification algorithm of the supervised classifier, a more principled solution is provided by performing the new physics search in a model-independent mode where, instead of trying to provide an experimental validation of a particular beyond the Standard Model physics model, one is simply looking for deviations from known Standard Model physics. The goal of model-independent new physics searches is to be sensitive to any deviations from known physics, whether or not they are described by one of the existing theoretical models. 
This kind of ideas have been put forward both at the Tevatron and at the LHC. For example, at the CDF experiment, a combination of algorithms called Vista and Sleuth were used to perform a global model-independent search of new physics with early Tevatron Run II data \cite{CDF2008}, while an algorithm called MUSiC is currently being used to scan the data of the CMS experiment at the LHC for such deviations \cite{CMS2008}.

In this paper, we present an alternative algorithm for model-independent searches of new physics based on semi-supervised anomaly detection techniques. As opposed to existing methods, our aim is not only to detect deviations from known Standard Model physics in a model-independent manner but also to produce a model for the observations which can be used to further analyze the properties of the potential new physics signal. Our method is also inherently multivariate and binning-free, while the existing methods are based on exploiting the properties of one-dimensional histograms. 

\section{Semi-supervised detection of collective anomalies}

\subsection{Motivation}

Our data analysis problem can be formulated as follows: given a labeled sample of known physics expected to be seen in the experiment and the empirical observations, how can we find out if there is a previously unknown contribution among the measurements and, if there is, how can we study its properties? In statistics, such a problem is solved using semi-supervised anomaly detection algorithms \cite{Chandola2009}. These are algorithms designed to look for deviations from a labeled sample of normal data. The neural network based background encapsulator \cite{Ribarics1993} of the trigger of the H1 experiment at HERA can, for instance, be regarded as an example of such an algorithm.

Unfortunately, existing semi-supervised anomaly detection algorithms can rarely be directly applied to solve the model-independent search problem. This is because they are designed to classify observations as anomalies should they fall in regions of the data space where there is a small density of normal observations. The problem is that in many theoretical models the excess of events corresponding to the new physics signal appears among the bulk of the background distribution instead of the tail regions considered by standard anomaly detection algorithms. We solve the problem by exploiting the fact that if there is a new physics signal, it should occur as a collection of several events which, only when considered together, constitute an anomalous deviation from the Standard Model.

\subsection{Fixed-background model for anomaly detection}

We start by fitting a parametric density estimate $p_\mathrm{B}(\bm{x})$ to a labeled background sample. This density estimate, which we call the \emph{background model}, serves as a reference of normal data. We conduct a search for collective deviations from the background model by fitting to the observations a mixture of $p_\mathrm{B}(\bm{x})$ and an additional \emph{anomaly model} $p_\mathrm{A}(\bm{x})$:
\begin{equation} \label{eq:pFB}
p_{\mathrm{FB}}(\bm{x}) = (1-\lambda)p_{\mathrm{B}}(\bm{x}) + \lambda
p_{\mathrm{A}}(\bm{x}).
\end{equation}
We call this mixture model the \emph{fixed-background model} to emphasize that when modeling the observations, $p_\mathrm{B}(\bm{x})$ is kept fixed which allows $p_\mathrm{A}(\bm{x})$ to capture any unmodeled anomalous contributions in the data. Fitting of both $p_\mathrm{B}(\bm{x})$ and $p_{\mathrm{FB}}(\bm{x})$ to the data is done by maximizing the likelihood of the model parameters.

Pattern recognition of the anomalies with the fixed-background model enables a variety of data analysis tasks:
\begin{enumerate}
 \item \emph{Classification}: Observations can be classified as anomalies using the posterior probability as a discriminant function
\begin{equation} \label{eq:d}
p(\mathrm{anomaly}|\bm{x}) = \frac{\lambda
p_{\mathrm{A}}(\bm{x})}{(1-\lambda)p_{\mathrm{B}}(\bm{x}) + \lambda
p_{\mathrm{A}}(\bm{x})} =: \mathcal{D}(\bm{x}).
\end{equation}
An observation $\bm{x}$ is then classified as an anomaly if $\mathcal{D}(\bm{x}) \geq T$ for some threshold $T \in [0,1]$.
 \item \emph{Proportion of anomalies}: The mixing proportion of the anomaly
model $\lambda$
directly gives us an estimate for the proportion of anomalies in the observations. This could be further used to derive a cross section estimate for the anomalous physics process.
 \item \emph{Statistical significance}: The statistical significance of the
anomaly model
can be estimated by performing a statistical test for the background-only null
hypothesis $\lambda = 0$ using the likelihood ratio test statistic $\Lambda$ \cite{Knight2000}. This enables us to discriminate between statistical
fluctuations of the background and real anomalous processes. Following
\cite{Wang1997},
we obtain the distribution of the test statistic
using nonparametric bootstrapping. That is, we sample with
replacement
observations from the background data, fit $p_{\mathrm{FB}}(\bm{x})$ to this new
sample
and compute the corresponding value of $\Lambda$. This allows us to recover the distribution of $\Lambda$ under the
background-only null
hypothesis and hence to
compute the significance of the observed collective anomaly.
\end{enumerate}

\begin{figure}[tb]
\centering
\subfigure{
\includegraphics[width=7cm]{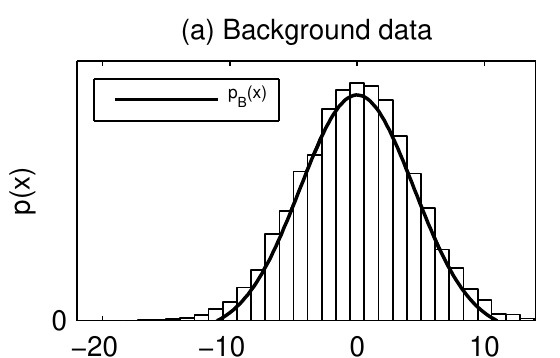}
}~
\subfigure{
\includegraphics[width=7cm]{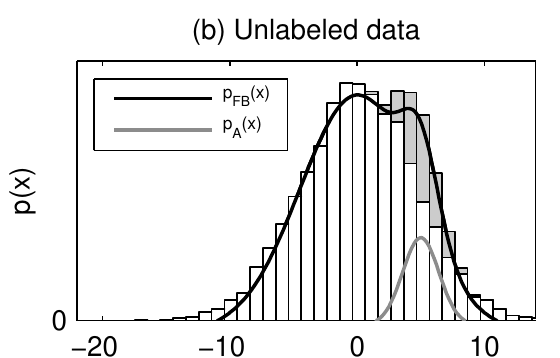}
}
\caption{(a) A histogram of background data from a univariate Gaussian distribution and an estimated background model $p_{\mathrm{B}}(x)$. (b) An
illustration of the fixed-background model in a univariate case. The histogram shows the unlabeled observations (the gray excess in the histogram
denotes the anomalous contribution). The estimated fixed-background model $p_{\mathrm{FB}}(x)$ is shown with a black line and the anomaly model $p_{\mathrm{A}}(x)$ with a gray line.}
\label{fig:illustration}
\end{figure}

As a simple illustration of the fixed-background model, Figure~\ref{fig:illustration}a shows a univariate data set of background data 
generated from a Gaussian distribution and a maximum likelihood Gaussian density $p_{\mathrm{B}}(x)$ estimated using the data set. Figure~\ref{fig:illustration}b shows a very simple anomalous pattern that can be modeled with a single additional univariate Gaussian. Given a sample contaminated with these anomalies, our goal is to find an optimal combination of the parameters of the anomaly model ($\mu_\mathrm{A}$, $\sigma_\mathrm{A}^2$) and the mixing proportion $\lambda$. The resulting model $p_{\mathrm{FB}}(x)$ is shown with a black line and the anomaly model $p_{\mathrm{A}}(x)$ with a gray line in Figure~\ref{fig:illustration}b.

\subsection{Algorithmic details}

We model all the densities in Equation \eqref{eq:pFB} using a multivariate
Gaussian mixture model \cite{McLachlan2000}
\begin{equation}
 p(\bm{x}|\bm{\theta}) = \displaystyle\sum_{j=1}^J \pi_j
\mathcal{N}(\bm{x}|\bm{\mu}_j, \bm{\Sigma}_j).
\end{equation}
Here $J$ is the number of Gaussian components in the model. The components have means $\bm{\mu}_j$ and covariances $\bm{\Sigma}_j$, while $\pi_j$ are the mixing proportions which satisfy $\sum_j \pi_j = 1$.

Fitting a Gaussian mixture model with $J$ components to the background sample by maximizing its log-likelihood is a standard problem in computational statistics and can be carried out using the iterative expectation-maximization (EM) algorithm \cite{McLachlan2008}. The algorithm proceeds in two steps. In the \emph{expectation step} (E-step), we compute the posterior probabilities
for each data point $\bm{x}_i$ to have been generated by the $j$th Gaussian component
\begin{equation}
\label{estep}
p(z_{ij}=1|\bm{x}_i,\bm{\theta}^k) =
\frac{\pi_j^k\mathcal{N}(\bm{x}_i|\bm{\mu}_j^k,
\bm{\Sigma}_j^k)}{\sum_{j'=1}^J\pi_{j'}^k
\mathcal{N}(\bm{x}_i|\bm{\mu}_{j'}^k
\bm{\Sigma}_{j'}^k)} =: \gamma_{ij}^k.
\end{equation}
Here, $\bm{\theta}^k$ contains the parameter estimates at
the $k$th iteration and $\bm{z}_i$ indicates which component generated the
$i$th observation. In the subsequent \emph{maximization step} (M-step), the parameter values are
updated according to the following equations:
\begin{equation}
 \pi_j^{k+1} = \frac{1}{N} \sum_{i=1}^N \gamma_{ij}^k, \quad
 \bm{\mu}_j^{k+1} = \frac{\sum_{i=1}^N \gamma_{ij}^k \bm{x}_i}{\sum_{i=1}^N
\gamma_{ij}^k}, \quad
 \bm{\Sigma}_j^{k+1} = \frac{\sum_{i=1}^N
\gamma_{ij}^k (\bm{x}_i-\bm{\mu}_j^{k+1})(\bm{x}_i-\bm{\mu}_j^{k+1})^{
\textrm{T}
}}{\sum_{i=1}^N \gamma_{ij}^k}.
\end{equation}
The algorithm alternates between these two steps until the log-likelihood is not improving anymore. This gives us an estimate of the background model $p_\mathrm{B}(\bm{x})$.

The next step of the anomaly detection algorithm is to fit to the unlabeled observations the fixed-background model \eqref{eq:pFB} where the anomaly model is a Gaussian mixture model with $Q$ components. When we keep the background model fixed, the free parameters of the full model are the parameters of the anomalous Gaussians, their mixing proportions and the global mixing proportion $\lambda$ and we seek to find such values of these parameter that their log-likelihood is maximized. This can be done easily by noting that as a mixture of two Gaussian mixtures, the fixed-background model itself is a Gaussian mixture with $J+Q$ components. Hence, we can use the same EM update equations as above to update the free parameters of the model and simply skip the updates of the fixed parameters. The complete update equations based on this idea can be found in the technical report \cite{Vatanen2011}, where we also describe a number of heuristics implemented in the algorithm to overcome the issues related to choosing the model complexities $J$ and $Q$.


\section{Demonstration: Search for the Higgs boson}

We demonstrate the application potential of the proposed anomaly detection framework by using a data set from the Higgs boson analysis at the CDF detector. It should be stressed that the results presented here should not be regarded as a realistic Higgs analysis. Instead, the goal is to merely demonstrate the performance and the potential benefits of the proposed algorithm.

\subsection{Description of the data set}

We consider a data set produced by the CDF collaboration \cite{Nagai2009}
containing background events and MC simulated Higgs events where the
Higgs is produced in association with the $W$ boson and decays into two bottom
quarks, $q\bar{q} \rightarrow WH \rightarrow l \nu b \bar{b}$. In the data space, this signal looks
slightly different for different Higgs masses $m_\textrm{H}$. The goal is to show that semi-supervised
anomaly
detection is able to identify such a signal without a priori knowledge of
$m_\textrm{H}$.
More generally, this could be any set of free parameters in the new physics theory
under consideration.

Each observation in the data set corresponds to a single simulated collision
event in the CDF detector at the Tevatron proton-antiproton collider. We follow
the event selection and choice of variables described in \cite{Nagai2009} for double
SECVTX tagged collision events. We also consider an additional neural network based
flavor separator from \cite{Chwalek2007}, giving us a total of 8 variables to describe
each event. To facilitate density estimation and visualization of the results, the
dimensionality of the logarithmically normalized data was reduced to 2 using principal
component analysis (PCA) \cite{Jolliffe2002}.

We used 3406 data points to train the background model which was then used to
detect signals of 400 data points for masses $m_\textrm{H} = 100, 115, 135,
150$~GeV among another sample of 3406 observations of
background data. Hence, the unlabeled sample contained 10.5\:\% of signal
events. In reality, the expected signal is roughly 5 to 50 times weaker than
this,
but due to the limited number of background events available, the signal had to
be amplified for this demonstration. Based on experiments with artificial toy data
reported in \cite{Vatanen2011}, we expect to be able to detect signals which contribute
only a few percent to the unlabeled sample when we have two orders of magnitude larger
background statistics.

\subsection{Modeling the Higgs data}

\begin{figure}[tb]
\centering
\subfigure{
\includegraphics[width=6cm]{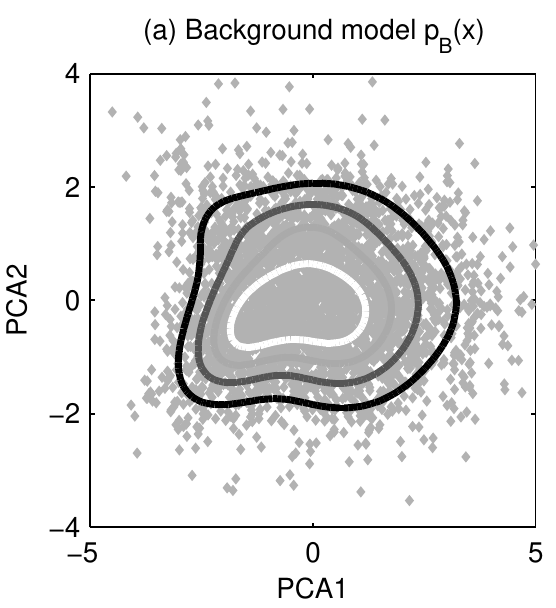}
}
\hspace{1cm}
\subfigure{
\includegraphics[width=6cm]{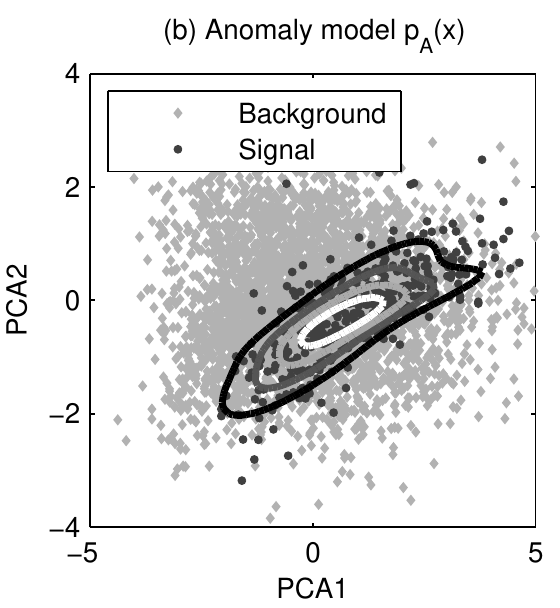}
}
\caption{(a) A projection of the WH background data into its
two-dimensional principal subspace. The solid lines show contours of the
estimated 5-component Gaussian mixture model for the background. (b) A projection of the
$m_\textrm{H} =
150$~GeV test data set into the two-dimensional principal subspace.
The solid lines show contours of the estimated 2-component Gaussian mixture model for the signal.}
\label{fig:higgsModel}
\end{figure}

We used the cross-validation-based information criterion (CVIC) \cite{Smyth2000}
to select a suitable number of components $J$ for the background model.
When a 5-fold cross-validation was performed, the evaluation log-likelihood was
maximized with $J=5$. Figure~\ref{fig:higgsModel}a shows contours of the resulting
five-component background model in the two-dimensional principal subspace.

We then learned the fixed-background models for the signals with different
masses starting with $Q = 3$ anomalous components. The algorithm converged
with one anomalous component for $m_\textrm{H} = 100$~GeV and two components
for the rest of the masses. The resulting anomaly model for $m_\textrm{H} =
150$~GeV is shown in Figure~\ref{fig:higgsModel}b.

\subsection{Anomaly detection results}

The statistical significances of the anomaly models were evaluated using the
bootstrap technique with $50\:000$ resamplings. The significances, starting
from the lowest mass, were $1.8\sigma$, $2.8\sigma$, $3.1\sigma$ and $3.3\sigma$.
Hence, in our toy analysis, all the signals would have been significant enough
to draw closer attention.

\begin{figure}[tb]
\centering
\subfigure{
\includegraphics[width=7.2cm]{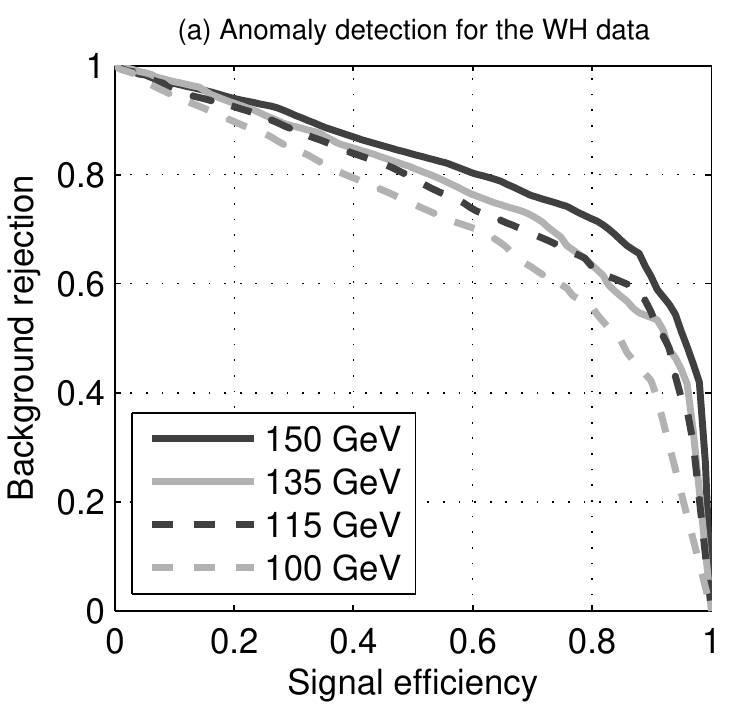}
}\hspace{0.5cm}
\subfigure{
\includegraphics[width=7.2cm]{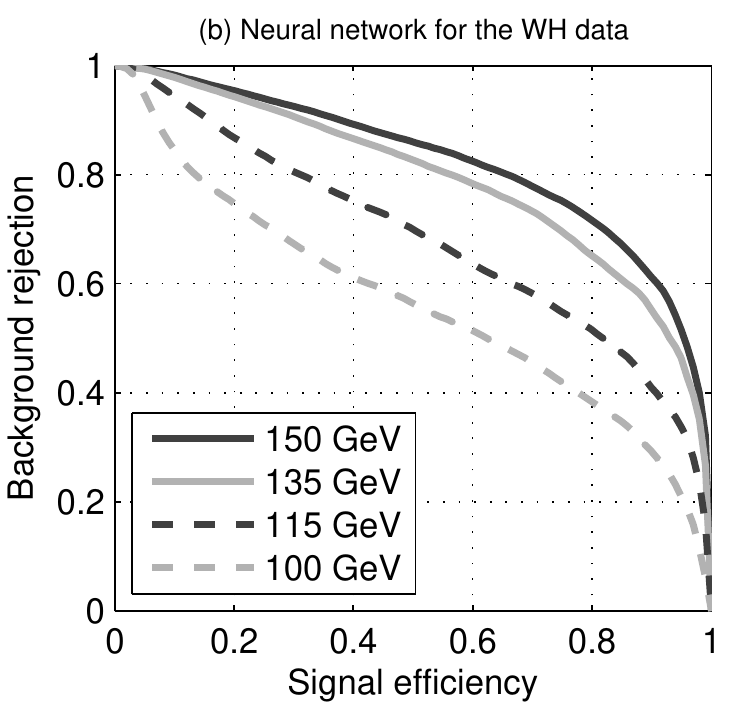}
}
\caption{(a)~ROC curves for the WH data
with different Higgs masses $m_\textrm{H}$ with semi-supervised anomaly detection.
The method is able to identify the signal without a priori knowledge of the mass.
(b)~ROC curves for the WH data for a neural network classifier trained with the 150~GeV signal.
The neural network is able to efficiently identify only the signal it has been trained for.}
\label{fig:higgsROC}
\end{figure}

Figure~\ref{fig:higgsROC}a shows the receiver operating characteristic
(ROC) curves for anomaly detection
with different Higgs masses. One can see that regardless of the mass
of the Higgs, the algorithm is able to identify the signal with
a relatively constant accuracy.
The classification results are slightly better with the
higher masses because the high-mass signal lies on a region of the data space
with a lower background density. Starting from
the lowest mass, the estimated anomaly proportions are $\lambda =
0.100,\:0.121,\:0.118,\:0.122$, which are all in agreement with the correct
proportion of 0.105.

We also trained a supervised MLP neural network for each of the mass points to
act as an ``optimal'' reference
classifier and compared the ROC curves to the ones obtained with anomaly
detection. Figure \ref{fig:higgsROC}b shows the ROC curves for a neural network
trained with the 150~GeV signal and tested with all the mass points.
When the neural network was tested
and trained with the same mass, the ROC curve was comparable to anomaly detection, which shows that
the proposed model-independent framework is able to achieve similar performance as
model-dependent supervised classification. However, when the neural network was tested
with a mass different from its training mass, its performance started to decline.
For example the 150~GeV neural network would be likely to miss the 100~GeV signal.
On the other hand, anomaly detection is able to successfully identify the signal
regardless of the mass. In other words, this experiment shows that supervised
classifiers are able to efficiently identify only the signals they have been
trained for with potentially severe consequences, while model-independent
approaches, such as semi-supervised anomaly detection, do not suffer from such
a limitation.

\section{Discussion}

The semi-supervised anomaly detection algorithm could be used to scan the measurements for new physics signals
by focusing on some particular final state which is thought to be especially sensitive to new physics.
One could then use the framework to look for deviations from the expected Standard Model background in this final
state. If a statistically significant anomaly is found, one could use the fixed-background model to study its
properties. Since the anomalous events are likely to lie within the bulk
of the background, the best way to reconstruct their properties would be a soft
classification based approach \cite{Kuusela2011} where the contribution of a single event to some physical observable is
weighted by the posterior class probability of the event. Reconstructing a number of physical spectra for the anomaly
should allow one to produce a physics interpretation for the observed deviation. It is likely that
most observed anomalies correspond detector effects and background mismodeling. If this is determined
to be the case, new cuts could be introduced to isolate such regions or the background estimate could
be corrected to account for the anomaly. The analysis could
then be repeated iteratively until all anomalies are understood. If at some stage we encounter
a significant anomaly which cannot be explained by just adjusting the background estimate, it could
be a sign of new physics and should be studied further to see if there is a plausible new physics
interpretation for it.

The computational experiments of this paper were carried out using well-understood standard techniques from computational statistics in order to convey the basic idea of semi-supervised anomaly detection as clearly as possible. It is likely that some of the shortcomings of the algorithm could be alleviated by using more advanced computational tools. One of the obvious limitations of the algorithm is the curse of dimensionality.
With a reasonable sample size, the algorithm seems to perform relatively well up to three dimensions,
but beyond that, the number of observations required to estimate the parameters of the Gaussian mixture models becomes prohibitively large. We demonstrated with the Higgs example that one possible way of solving the problem is dimensionality reduction with PCA or some other dimensionality reduction algorithm. Another possibility would be to consider parsimonious Gaussian mixture models, where the number of parameters is reduced by constraining the structure of the covariance matrices \cite{Fraley2002}. The current algorithm
is also only able to handle anomalies which manifest themselves as an excess over the background. That is,
a deficit with respect to the background estimate
is not treated properly, although it might be possible to circumvent this restriction by allowing
$\lambda$ to take negative values.

\section{Conclusions}

We presented a novel and self-consistent framework for model-independent searches of new physics
based on a semi-supervised anomaly detection algorithm. We showed using a Higgs boson data set that
the method can be successfully applied to searches of new physics and demonstrated the potential
benefits of the approach with respect to conventional analyses.
To make sure that no new physics signals have been missed by the current model-dependent
searches, it would be important to complement
them by scanning the collision data with model-independent techniques, one example of which
is the proposed anomaly detection framework. We hope that the work presented here
helps to revive interest in such techniques among the HEP community by showing that
model-independent new physics searches can be conducted in a feasible and practical
manner.

\ack
The authors are grateful to the CDF collaboration for providing access to the Higgs signal and background Monte Carlo samples, to the Academy of Finland for financial support and to Matti P\"{o}ll\"{a}, Timo Honkela and Risto Orava for valuable advice.

\section*{References}
\bibliographystyle{iopart-num}
\bibliography{acat}

\end{document}